\documentclass[a4paper, 11pt, twoside]{article}
\usepackage[utf8x]{inputenc}
\usepackage{authblk} 
\usepackage{a4wide}  
\usepackage{graphicx} 
\usepackage{placeins} 
\usepackage{engord} 
\usepackage{enumitem}

\title{$\omega$ and $\eta$ meson production in p+p reactions at E$_{kin}$=~3.5~GeV}
\date{}
\author{K.~Teilab$^{7}$, G.~Agakishiev$^{6}$, A.~Balanda$^{3}$, D.~Belver$^{16}$, A.~Belyaev$^{6}$,
A.~Blanco$^{2}$, M.~B\"{o}hmer$^{12}$, J.~L.~Boyard$^{14}$, P.~Cabanelas$^{16}$, E.~Castro$^{16}$
% ...
}
\affil{\small
(HADES collaboration) \\
{$^{1}$Istituto Nazionale di Fisica Nucleare - Laboratori Nazionali del Sud, 95125~Catania, Italy}\\
{$^{2}$LIP-Laborat\'{o}rio de Instrumenta\c{c}\~{a}o e F\'{\i}sica Experimental de Part\'{\i}culas , 3004-516~Coimbra, Portugal}\\
{$^{3}$Smoluchowski Institute of Physics, Jagiellonian University of Cracow, 30-059~Krak\'{o}w, Poland}\\
{$^{4}$GSI Helmholtzzentrum f\"{u}r Schwerionenforschung GmbH, 64291~Darmstadt, Germany}\\
{$^{5}$Institut f\"{u}r Strahlenphysik, Forschungszentrum Dresden-Rossendorf, 01314~Dresden, Germany}\\
{$^{6}$Joint Institute of Nuclear Research, 141980~Dubna, Russia}\\
{$^{7}$Institut f\"{u}r Kernphysik, Goethe-Universit\"{a}t, 60438 ~Frankfurt, Germany}\\
{$^{8}$II.Physikalisches Institut, Justus Liebig Universit\"{a}t Giessen, 35392~Giessen, Germany}\\
{$^{9}$Istituto Nazionale di Fisica Nucleare, Sezione di Milano, 20133~Milano, Italy}\\
{$^{10}$Institute for Nuclear Research, Russian Academy of Science, 117312~Moscow, Russia}\\
{$^{11}$Excellence Cluster 'Origin and Structure of the Universe' , 85478~Munich, Germany}\\
{$^{12}$Physik Department E12, Technische Universit\"{a}t M\"{u}nchen, 85748~M\"{u}nchen, Germany}\\
{$^{13}$Department of Physics, University of Cyprus, 1678~Nicosia, Cyprus}\\
{$^{14}$Institut de Physique Nucl\'{e}aire (UMR 8608), CNRS/IN2P3 - Universit\'{e} Paris Sud, F-91406~Orsay Cedex, France}\\
{$^{15}$Nuclear Physics Institute, Academy of Sciences of Czech Republic, 25068~Rez, Czech Republic}\\
{$^{16}$Departamento de F\'{\i}sica de Part\'{\i}culas, Univ. de Santiago de Compostela, 15706~Santiago de Compostela, Spain}\\
{$^{17}$Instituto de F\'{\i}sica Corpuscular, Universidad de Valencia-CSIC, 46971~Valencia, Spain}\\
\vspace{\lineskip}
{$^{a}$ also at ISEC Coimbra, ~Coimbra, Portugal}\\
{$^{b}$ also at Technische Universit\"{a}t Dresden, 01062~Dresden, Germany}\\
{$^{c}$ also at Dipartimento di Fisica, Universit\`{a} di Milano, 20133~Milano, Italy}\\
{$^{d}$ also at Panstwowa Wyzsza Szkola Zawodowa , 33-300~Nowy Sacz, Poland}\\
{$^{e}$ also at Dipartimento di Fisica Generale, Universita' di Torino, 10125~Torino, Italy}}

\begin{document}
\thispagestyle{empty}
{\renewcommand{\thefootnote}{} \footnotetext{Presented at the
\engordnumber{11} International Workshop on Meson Production, Properties
and Interaction, KRAK\'{O}W, POLAND, 10 - 15 June 2010}}
\maketitle

\begin{abstract}
\noindent
We report on the exclusive production of $\omega$ and $\eta$ mesons
in $p+p$ reactions at 3.5 GeV beam kinetic energy.  Production cross
sections, angular distributions and Dalitz plots of both mesons were determined.
Moreover, the relative contribution of the $N$(1535) resonance in $\eta$
production at this energy was evaluated.
\newline
We conclude that $\eta$ mesons
produced via $N$(1535) exihibit an isotropic angular distribution,
whereas those produced directly show a strong anisotropic distribition.
$\omega$ mesons show a slightly anisotropic angular distribition.

\end{abstract}

\section{Introduction and Motivation}

The determination of integrated and differential cross sections
for the production of light mesons in the energy range up to
1~GeV above threshold in proton-proton reactions is very important
for the understanding of the underlying elementary mechanisms as well
as for the interpretation of heavy ion results. However, most of the
high-precision and high-statistics experiments were performed near
threshold. At higher energies, a large number of older data sets are
available but only for integrated cross sections\cite{flam}.

Concerning the $\eta$ production at such energies, the DISTO
collaboration provided a comprehensive set of measurements,
including momentum and angular distributions as well as
$pp\eta$ Dalitz plots, all at three different beam energies
(2.15, 2.5 and 2.85 GeV)\cite{bale2}. 
However, the provided angular distributions were integrated
over the whole Dalitz plot.
\newline
Moreover, the DISTO collaboration studied the angular distribution
of $\omega$ mesons in $p+p$ reactions at 2.85~GeV beam kinetic
energy\cite{bale1}. 

In this work, we present new measurements of integrated and
differential cross sections for the production of $\eta$ and $\omega$
mesons in $p+p$ reactions at 3.5~GeV beam kinetic energy using
the magnetic spectrometer HADES\cite{nim} (\textbf{H}igh
\textbf{A}cceptance \textbf{D}i-\textbf{E}lectron \textbf{S}pectrometer).

\section{Analysis}

The reconstruction of $\omega$ and $\eta$ mesons was done via
their decay into three pions ($\pi^+\pi^-\pi^0$). The two protons
and charged pions were identified using energy loss information in
the drift chambers. The $\pi^0$ was reconstructed via the missing
mass method. To improve the momentum resolution, a kinematical fit
was applied. The background channels were removed by a cut on the
probability of the fit. The $\omega$ and $\eta$ mesons can be clearly
identified as two peaks on top of a continous background in the two 
protons missing mass spectrum. 

\section{Results}

\subsection{$\omega$ meson}

\begin{figure}[ht]
  \centering
  a)
  \includegraphics[width=7cm]{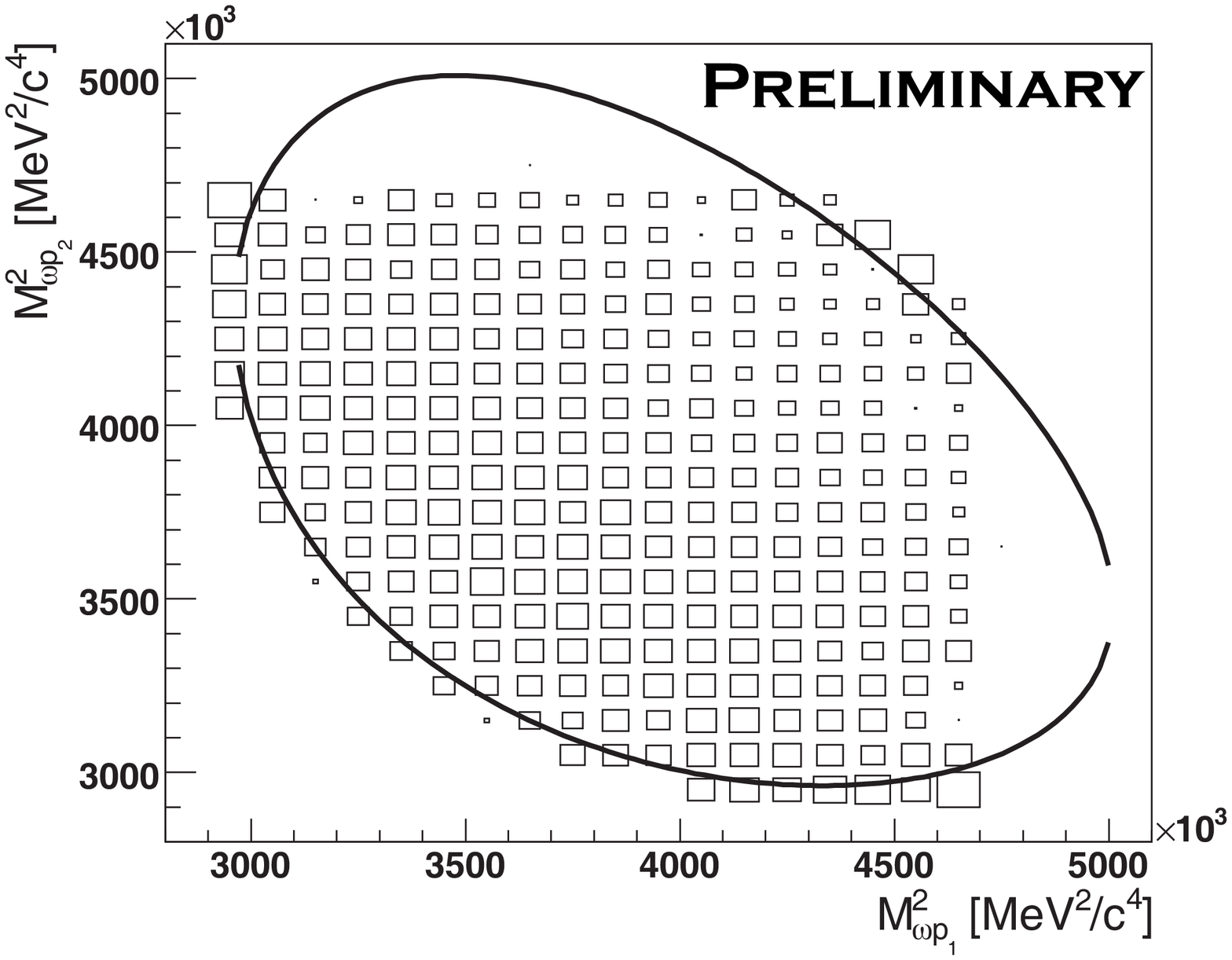} 
  b) 
  \includegraphics[width=7cm]{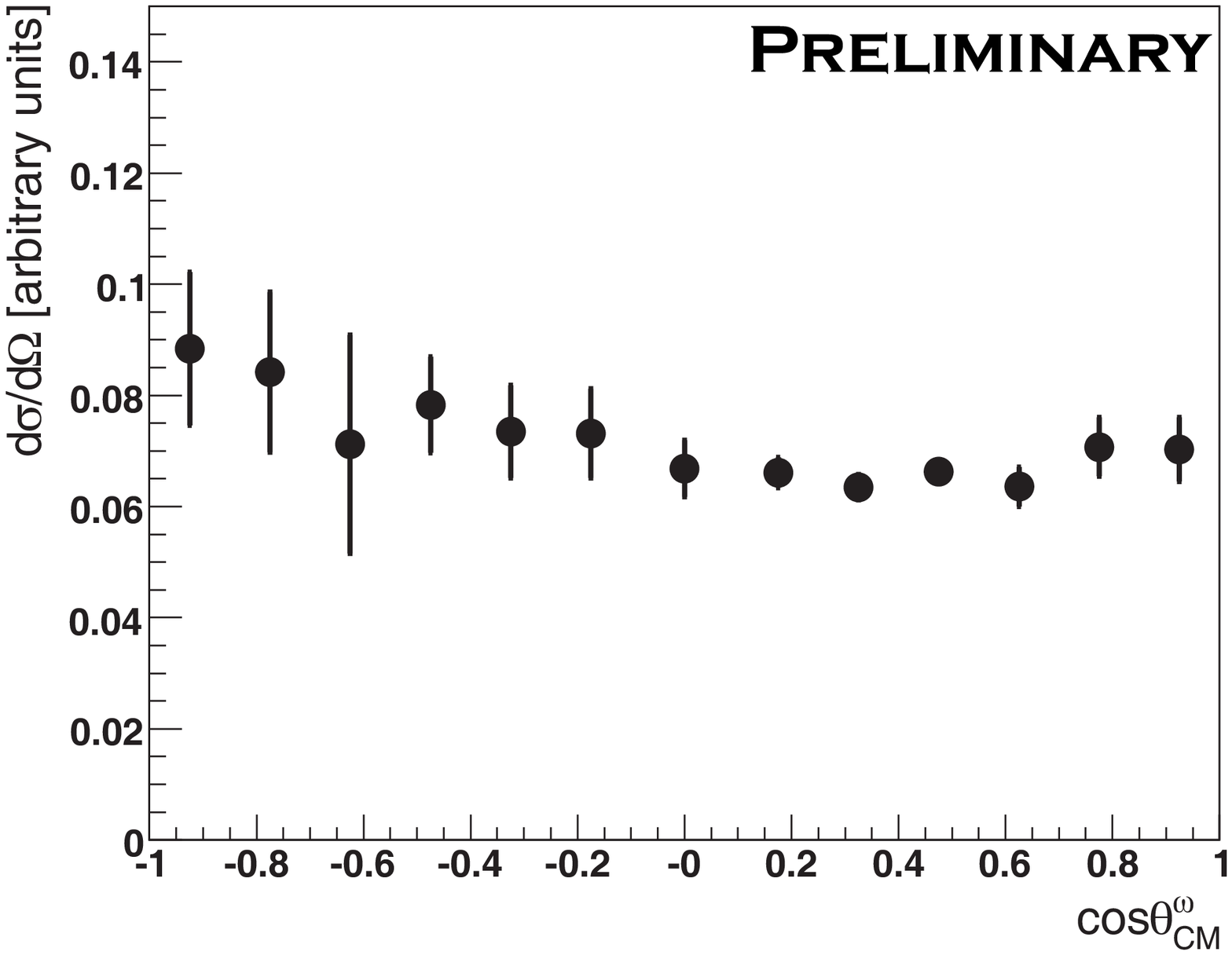}
  \caption{a) $pp\omega$ Dalitz plot. The thick line shows the phase space
           limit. b) Polar angular distribution of $\omega$ emission in the c.m. system.
           Both plots are corrected for acceptance. Empty bins correspond to zero acceptance.} 
\label{fig:omega}
\end{figure}

For the $\omega$ meson production, we present in fig. \ref{fig:omega} the
$pp\omega$ Dalitz plot and polar angular distribution in the c.m. system.
All plots were corrected for the detector acceptance bin-by-bin using a
corresponding factor extracted from a PLUTO\cite{pluto} + GEANT simulation. 
\newline
We do not observe a strong contribution of resonances in the production
and we find that the polar angular distribution of $\omega$ mesons is only
slightly anisotropic. The determined cross section amounts to
(137$\pm$27)~$\mu$b (prelim.).

\subsection{$\eta$ meson}

\begin{figure}[ht]
  \centering 
  a)
  \includegraphics[width=7cm]{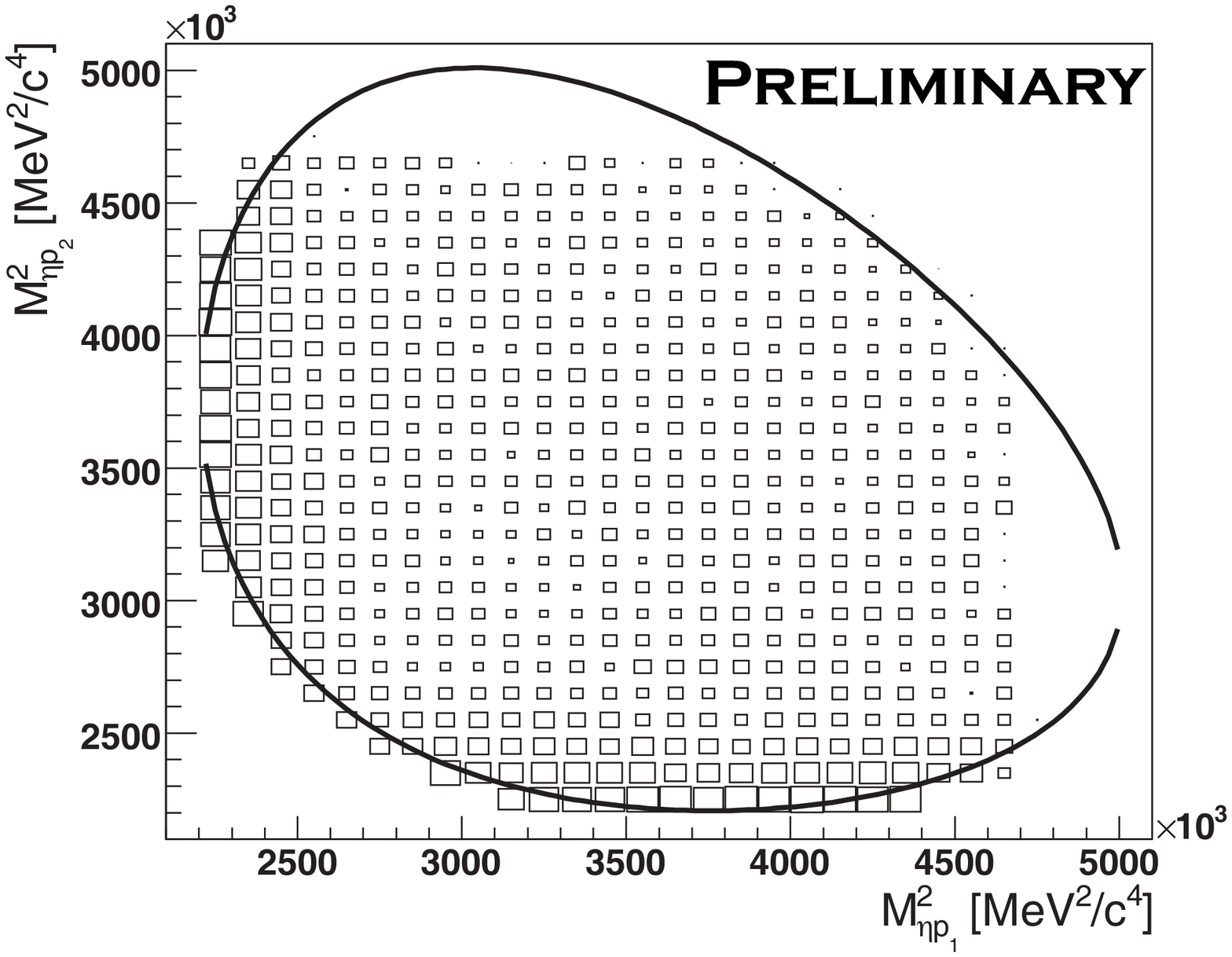}
  b) 
  \includegraphics[height=5.5cm]{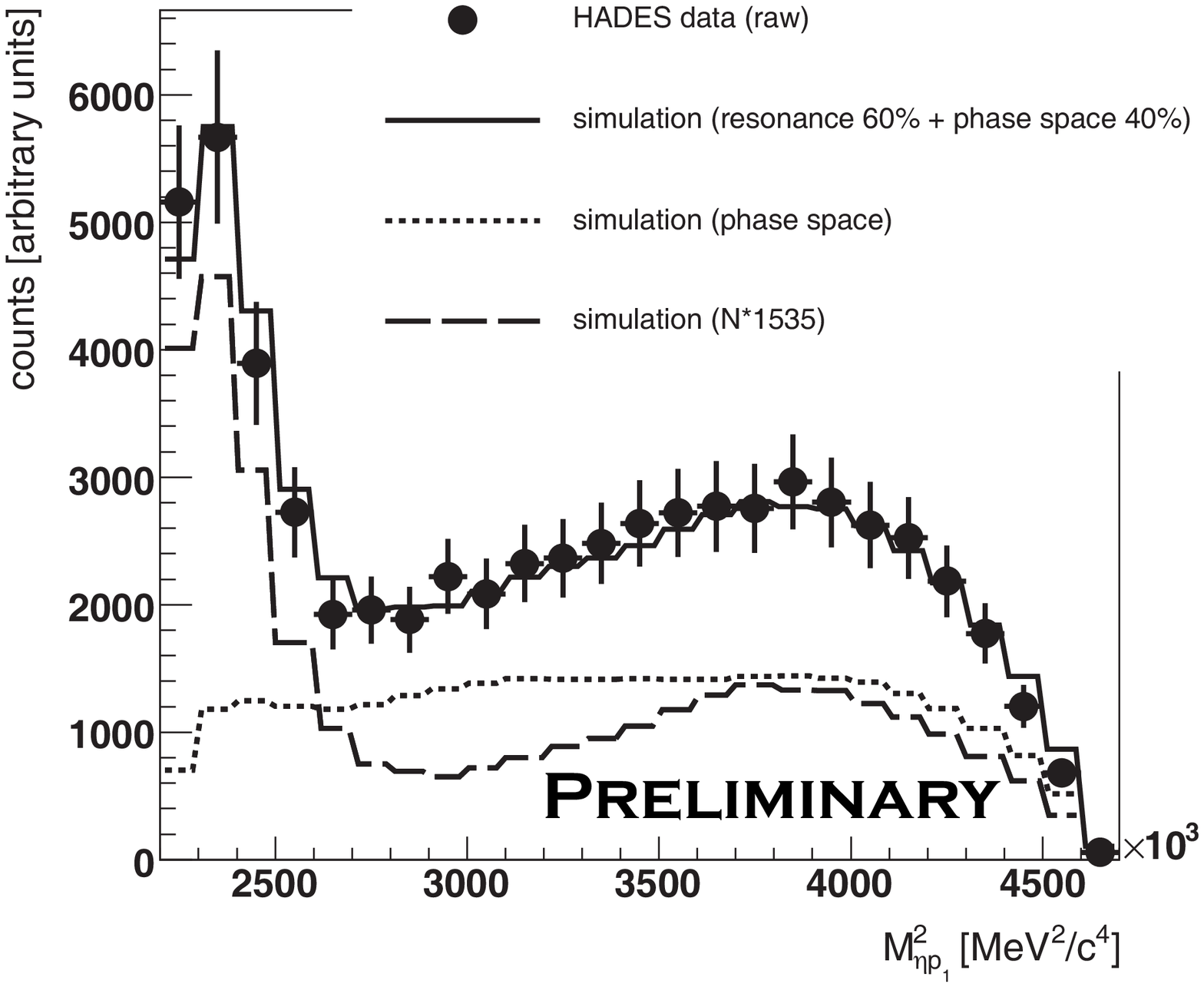}
  \caption{a) $pp\eta$ Dalitz plot. The thick line shows the phase space 
	   limit. The data are corrected for acceptance. Empty bins correspond
	   to zero acceptance. b) Projection of the raw Dalitz plot. The dotted
	   line shows the mass distribution as given by phase space. The dashed
	   line shows simulation of the mass distribution for production via $N$(1535).
}\label{fig:eta1}
\end{figure}

\noindent
The $pp\eta$ Dalitz plot (fig. \ref{fig:eta1}a) shows a clear signal from the
$N$(1535) resonance. To extract the relative contribution of the resonance
to the production we simulated the $p\eta$ mass distribution in two scenarios:
\begin{enumerate}[label=\alph{*})]
 \item $\eta$ production according to phase space.
 \item $\eta$ production via $N$(1535) resonance.
\end{enumerate}
The resulting spectra were scaled to fit the measured data (see fig.
\ref{fig:eta1}b). The relative contribution of $N$(1535) to the production was determined
to be about 41\% (prelim.).

\begin{figure}[ht]
  \centering
  a)
  \includegraphics[width=7cm]{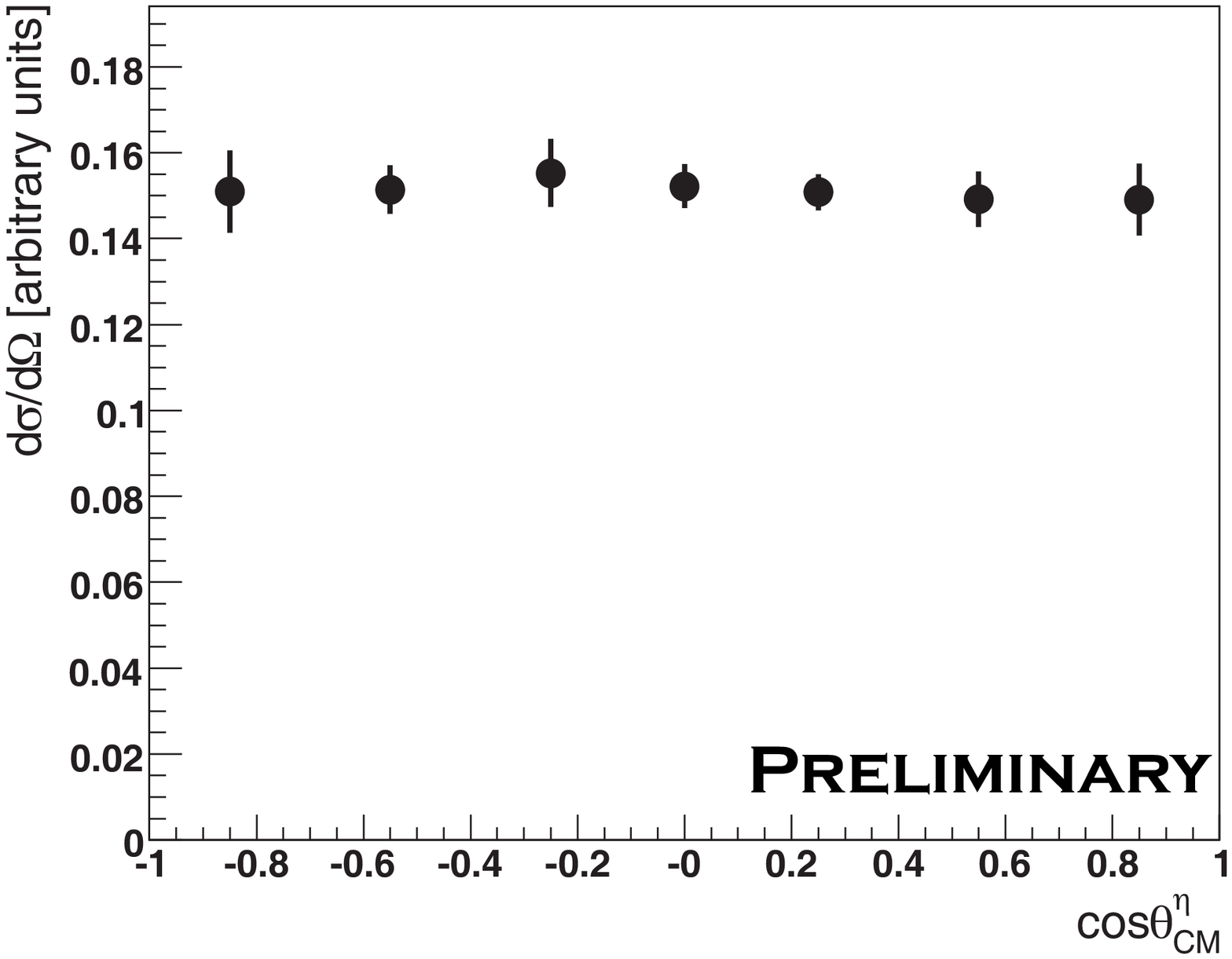}
  b)
  \includegraphics[width=7cm]{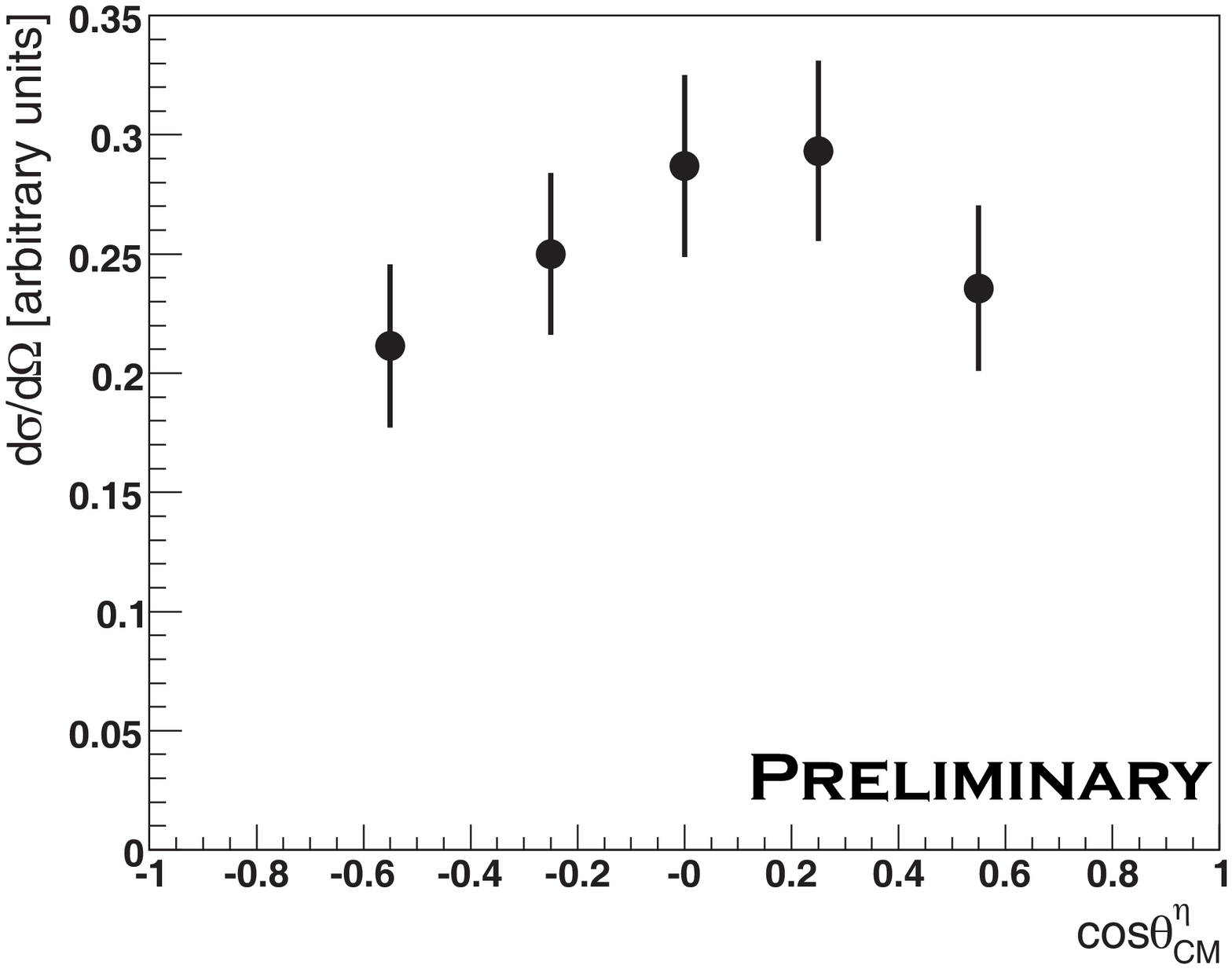}
  \caption{a) Polar angular distribution of $\eta$ emission in the c.m. system
	  (from resonance region). b) Polar angular distribution of $\eta$ emission
	  in the c.m. system (from non-resonant region). Both spectra are corrected
	  for acceptance. Empty bins correspond to zero acceptance.}\label{fig:eta}
\end{figure}

\noindent
The angular distribution of $\eta$ emission was determined for two regions of the
Dalitz plot: 
\begin{enumerate}[label=\Roman{*})]
\item the region where $M^2_{\eta+p_1}$ or $M^2_{\eta+p_2}$ $<$ 2.8 GeV$^2$/c$^4$.
      The $\eta$ production in this region is dominated by resonant production.
\item the region where $M^2_{\eta+p_1}$ and $M^2_{\eta+p_2}$ $>$ 2.8 GeV$^2$/c$^4$. The $\eta$
      production in this region is dominated by non-resonant production.
\end{enumerate}
\noindent
The angular distribution for region I is clearly flat as shown in fig. \ref{fig:eta}a,
whereas that of region II is strongly anisotropic (fig. \ref{fig:eta}b).
The total production cross section was determined to be (211$\pm$47)~$\mu$b (prelim.).

\section{Acknowledgments}
The HADES collaboration gratefully acknowledges the support
by BMBF grant 06MT9156, 06GI146I, 06FY171 and 06DR5059 (Germany),
by GSI (TM-FR1, GI/ME3, OF/STR),
by Excellence Cluster Universe (Germany),
by grants GA AS CR IAA100480803 and MSMT LC 07050 (Czech Republic),
by grant KBN 5P03B 140 20 (Poland),
by CNRS/IN2P3 (France),
by grants MCYT FPA2000-2041-C02-02
and XUGA PGID T02PXIC20605PN (Spain),
by INFN (Italy),
by grant UCY-10.3.11.12 (Cyprus),
by INTAS grant 06-1000012-8861
and EU contract RII3-CT-506078.

\end{document}